\documentclass{nature}
\usepackage{bbm}
\usepackage{graphicx}
\usepackage{dcolumn}
\usepackage{bm}
\usepackage{subfigure}
\usepackage{amsmath}
\usepackage{feynmf}
\usepackage{hyperref}
\usepackage{color}
\usepackage{changes}
\usepackage{ulem}

\usepackage{attachfile}

\usepackage[textwidth=14.5cm]{geometry}
\usepackage{blindtext}
\title{Higher-order Oscillatory Planar Hall Effect in Topological Kagome Metal}

\author{Leyi Li$^{1,3,\ast}$, Enkui Yi$^{1,3,\ast}$, Bin Wang$^{1}$, Guoqiang Yu$^{2,3}$, Bing Shen$^{1,3,\dagger}$, Zhongbo Yan$^{1,\ddagger}$ \& Meng Wang$^{1,\wr}$}

\begin{document}
\begin{sloppypar}

\maketitle

\begin{affiliations}
 \item Center for Neutron Science and Technology, Guangdong Provincial Key Laboratory of Magnetoelectric Physics and Devices, School of Physics, Sun Yat-Sen University, Guangzhou, Guangdong 510275, China
 \item Sate Key Laboratory of Optoelectronic Materials and Technologies, Sun Yat-Sen University, Guangzhou, Guangdong 510275, China
 \item Beijing National Laboratory for Condensed Matter Physics, Institute of Physics, University of Chinese Academy of Sciences, Chinese Academy of Sciences, Beijing 100190, China
\end{affiliations}

\begin{abstract}
Exploration of exotic transport behavior for quantum materials is of great interest and importance for revealing exotic orders to bring new physics. In this Letter, we report the observation of exotic prominent planar Hall effect (PHE) and planar anisotropic magnetoresistivity (PAMR) in strange kagome metal KV$_3$Sb$_5$. The PHE and PAMR, which are driven by an in-plane magnetic field and display sharp difference from other Hall effects driven by an out-of-plane magnetic field or magnetization, exhibit exotic higher-order oscillations in sharp contrast to those following empirical rule only allowing twofold symmetrical oscillations.  These higher-order oscillations exhibit strong field and temperature dependence and vanish around charge density wave (CDW) transition. The unique transport properties suggest a significant interplay of the lattice, magnetic and electronic structure in KV$_3$Sb$_5$. This interplay can couple the hidden anisotropy and transport electrons leading to the novel PHE and PAMR in contrast to other materials.
\end{abstract}

\textbf{INTRODUCTION}
\\Planar Hall effect (PHE) is a unique transport phenomena driven by an in-plane magnetic-field-induced rotation of the principal axes of the resistivity tensor\cite{KMKoch,CGoldberg}.
Because of different origins, the PHE is in many aspects quite different from those Hall effects driven by an out-of-plane magnetic field or magnetization as shown in Fig.1 (a). Obvious PHE has been observed in a few ferromagnetic metals\cite{KMKoch,CGoldberg} and nonmagnetic semimetals with strong orbital anisotropy of electronic structure\cite{QLiu,SYYang}. It is widely used for designing and fabricating commercial magnetic sensors, especially for three-dimensional (3D)  highly compacted, and ultra-sensitive 'lab-on-a-chip (LOC)' devices for the next generation chips. Another great interest for PHE lies in its angular dependence of the direction of magnetic field, which can be applied to infer the information of the underlying magnetic order or electronic structure. For example, to investigate this effect and pursue its origin in a quantum system may reveal  nontrivial topological physics or reveal the exotic states and orders to advance the understanding of fundamental physics
\cite{AABurkov,SNandy,HLi,FCChen,NKumar,AATaskin}.

Recently, superconductivity was observed in a new family of layered kagome metals AV$_3$Sb$_5$ (A = K, Rb, Cs)\cite{Gomes,Brenden, QiangweiYin,KYChen,FengDu,ZhuyiZhang}. Their normal states are identified as Z$_2$ topological metals with multiple topologically nontrivial band structures such as flat band, van Hove singularity and Dirac-fermion dispersion in close proximity to the Fermi level\cite{BROrtiz,MaximilianLKiesel,WSWang,XFeng,MMDenner,HaoxiangLi}. In the absence of magnetism, it is surprising to observe a giant anomalous Hall effect (AHE) in these materials. To reconcile
the observations,  the AHE is considered to  strongly correlate to novel orders such as chiral charge-density-wave (CDW) or nematic order accompanying with symmetry breaking\cite{YuXiaoJiang,ZhiweiWang,HeZhao,HuiChen,ShuoYingYang,HongLi,FHYu,EricMKenney,liyu}. However,
because the correlations among spin, charge, lattice, and  other orders are expected to be important,
it remains great challenge to get a full understanding of the origin of the various novel transport properties observed in this family of materials. In this work, we observe prominent PHE and planar anisotropic magnetoresistivity (PAMR) in KV$_3$Sb$_5$. More interestingly, below the CDW transition the in-plane applied magnetic field  drives exotic higher-order oscillations for PHE and PAMR violating the empirical law in former materials (even in recent topological materials). The unique behaviors exhibit strong field response accompanying with non-monotonous anisotropic field dependent in-plane resistivity. These novel planar transport properties suggest a strong coupling between transport electrons and anionotropies from the system with various orders in contrast to other materials. By scrutinizing various possible mechanisms, we suggest a significant interplay of the lattice, magnetism and electron associated with the fluctuations as the media leading to those novel transport behavior.

\textbf{RESULTS}
\\As shown in Fig.1 (a), the PHE and PAMR can be characterized by measuring the transverse resistivity ($\rho_{xy}$) and longitudinal resistivity ($\rho_{xx}$) with applying an in-plane magnetic field (rotating within the $ab$ plane of KV$_3$Sb$_5$ shown in Fig.1 (b)) which fails to drive the ordinary Hall effect (OHE) measured with an out-of-plane magnetic field. Usually, obvious PHE and PAMR are only observed in a few kinds of materials and  follow the  angular dependence as:
\begin{eqnarray}
\rho_{xy}&=&-\Delta \rho \sin \theta \cos \theta,\\
\rho_{xx}&=&\rho_{\perp} -\Delta \rho \cos ^2 \theta,
\end{eqnarray}
where $\rho_{xy}$ represents the in-plane Hall resistivity that directly shows the PHE, $\rho_{xx}$ is the PAMR, and $\Delta \rho=\rho_{\perp}-\rho_{\|}$ is the resistivity anisotropy (called chiral resistivity in topological materials) with  $\rho_{\perp}$ and $\rho_{\|}$ representing the resistivity with the applied field $\mu_0H$ perpendicular (90$^{\circ}$) and parallel (0$^{\circ}$) to the electric current respectively\cite{KMKoch}. According to these formulas, angular dependent $\rho_{xy}$ ($\rho_{xy} (\theta)$) and $\rho_{xx}$ ($\rho_{xx} (\theta)$) exhibit twofold oscillations and a relative 45$^{\circ}$-angle shift for $\rho_{xy} (\theta)$ and $\rho_{xx} (\theta)$ as shown in Figs.1(d) and (e)\cite{ShYY, Qiunan, ChenFC, Weib, YangXC, Gatti}. It is observed the anisotropy from lattice, magnetism, or Fermi surface etc usually hardly affect the in-plane transport behavior such as PHE and PAMR. For KV$_3$Sb$_5$, obvious PHE and PAMR are observed at 2 K  even with a small applied field of $\mu_0 H$ = 2 T after subtracting the OHE and out-of-plane magneto resistivity \cite{QiangweiYin,NKumar}. $\rho_{xy}$ and $\rho_{xx}$ (shown in Figs.1 (d) and (e) exhibit two-fold symmetrical oscillations and roughly follow the conventional angular dependence. The little discrepancy between our results and empirical formulas can be attributed to the existence of additional higher-order oscillations (discussed below).

With increasing the applied field ($\mu_0 H$), as shown in Fig. 2, at 2 K  additional features are observed in $\rho_{xy}$  at 90$^{\circ}$ and 270$^{\circ}$  with $\mu_0 H$= 7 T accompanying with the emergence of additional small peaks in $\rho_{xx} (\theta)$ at 0$^{\circ}$ and 180$^{\circ}$. With further increasing $\mu_0H$ to 14 T, the oscillations of both $\rho_{xy} (\theta)$ and $\rho_{xx} (\theta)$ exhibit prominent higher-order oscillatory components in sharp contrast to those for conventional PHE and PAMR with only two-fold symmetrical oscillations. To investigate these phenomenons, Fast Fourier Transform (FFT) is applied for $\rho_{xy} (\theta)$ and $\rho_{xx} (\theta)$ at 2 K with various applied fields as shown in Figs. 3 (a) and (b). Besides the peaks at 180$^{\circ}$ in FFT spectrum relating to two-fold symmetrical oscillations, another obvious peaks at 90$^{\circ}$ are observed at high fields whose amplitude is larger than those at 180$^{\circ}$, indicating the appearance of four-fold  symmetrical oscillations. With decreasing $\mu_0 H$,  these four-fold symmetrical oscillations become weaker and their amplitudes shrink to the value smaller than those with two-fold symmetry at $\mu_0 H$ = 5 T shown in Figs. 3 (c) and (d). With decreasing $\mu_0 H$ to 2 T,  the FFT peaks at 90$^{\circ}$ are much weaker than those at 180$^{\circ}$, indicating the oscillations mainly containing two-fold symmetry component.  In addition, small peaks are observed at 60$^{\circ}$ in the FFT spectrum for $\rho_{xx} (\theta)$, suggesting the appearance of six-fold symmetrical oscillation which are much weaker than the two-fold and four-fold symmetrical oscillations. To gain more clues on the emergence of higher-order oscillatory components, the field dependent Hall and longitudinal resistivity $\rho_{xy} (H)$ and $\rho_{xx} (H)$ are measured with rotating an in-plane applied field at angles from 0$^{\circ}$ to 90$^{\circ}$ shown in Fig. 3 (e) and (f). In these colour contour plots, it is clearly observed that minimums appear around 45$^{\circ}$ and 20$^{\circ}$ for $\rho_{xy} (H)$ and $\rho_{xx} (H)$ at high fields, respectively, which is consistent with the emergence of additional higher-order oscillatory components (especially the four-fold symmetrical oscillatory components ) in our former results and analysis for $\rho_{xy} (\theta)$ and $\rho_{xx} (\theta)$.

Figs. 4 (a) and (b) present the temperature dependence of $\rho_{xy} (\theta)$ and $\rho_{xx} (\theta)$ with $\mu_0 H$ = 14 T, respectively. By the similar FFT analysis shown in Figs. 4 (c) and (d), it is observed that  the high-order  oscillatory components relating to the four-fold symmetrical oscillation appear below the CDW transition (identified by an anomaly in temperature-dependent resistivity along $c$ axis around 80 K here), and increase obviously with decreasing the temperature, as shown in Fig. 4(e). Above 80 K, the oscillations of $\rho_{xy} (\theta)$ and $\rho_{xx} (\theta)$  only contain two-fold symmetrical oscillatory component and can be well described by  empirical formulas 1 and 2. To pursue the origin of the observed unconventional PHE and PAMR, we first carefully check the extrinsic contributions such as current jetting effect due to the inhomogeneous current distribution\cite{SLiang}. The possible tiny current distortion in our devices is difficult to drive additional prominent oscillations here (see the Supplemental Material).  The intrinsic origins usullay can be (1) the spin-dependent scattering due to the classical orbital magnetism or the interactions of the magnetic order and the spin-orbit coupling\cite{SLiang}, (2) chiral anomaly in a nonmagnetic topological system \cite{AABurkov}, and (3) anisotropically lifting the protection of surface Dirac fermions from backscattering in a topological insulator\cite{AATaskin}. However, these mechanisms can only host to the two-fold symmetrical oscillations for PHE and PAMR obeying the conventional formulas (1) and (2) and fail to result in high-order  oscillatory components.

\textbf{DISCUSSION}
\\Usually the additional anisotropy from a system is difficult to behave in conventional PHE and PAMR even the system hosts high anisotropic Fermi surface, magnetism, or lattice such as in kagome magnet Co$_3$Sn$_2$S$_2$, tetragonal ZrSiSe, and quasi-one dimensional TaSe$_3$\cite{ShYY, Qiunan, ChenFC, Weib, YangXC, Gatti}. However, these additional anisotropies should be considered in KV$_3$Sb$_5$. The anisotropic Fermi surfaces (triangular-shaped pockets) are observed around $M$ point\cite{ShuoYingYang} , but the shape of the Fermi surface changes little in the presence of CDW order and higher-orders oscillatory PHE and PAMR, which is unlikely to bring the appearance of large anisotropy in carrier density and scattering rate directly. Thus the mechanism applied to explain the higher-order oscillations in materials such as bismuth or (111) LaAlO$_{3}$/SrTiO$_{3}$ oxide interface\cite{PKRout,AYamada,SYYang} is  not applicable. In KV$_3$Sb$_5$ no long-range or short-range magnetic orders were observed, thus the anisotropy from conventional static magnetic order also seems not applicable. In semi-class model, the resistivity comes from the electron scattering with the lattice. Thus the anisotropy of the lattice can naturally host anisotropic carrier scattering with constrictions of the lattice's symmetry. It is noticed that even in presence of some distortion due to the CDW the lattice for KV$_3$Sb$_5$ still keeps hexagonal structure. In the two dimensional case with $C_{6}$ rotation invariance\cite{PKRout}, $\rho_{xy} (\theta)$ and $\rho_{xx} (\theta)$ can be expressed as:

\begin{eqnarray}
\rho_{xy}&=&S_{xy2}\sin(2\theta-2\phi)+S_{xy4}\sin(4\theta+2\phi),\\
\rho_{xx}&=&C_{xx0}+C_{xx2}\cos(2\theta-2\phi)+C_{xx4}\cos(4\theta+2\phi)\nonumber\\
&&+C_{xx6}\cos6\theta,
\end{eqnarray}

where S$_{xy2}$, C$_{xy2}$, S$_{xy4}$, C$_{xy4}$, and C$_{xy6}$, are the coefficients for two, four, and six -fold symmetrical oscillations for PHE and PAMR related to lattice symmetry\cite{sup,PKRout}.  $\phi$ is the angle between the current $I$ and $a$ axis and here $\phi$=0 with $I // a$. By using these formulas, the our data can be well fitted in consisted with former FFT analysis. Above CDW transition S$_{xy4}$, C$_{xy4}$, and C$_{xy6}$ become to zero, the formula 3 and 4 become to formula 1 and 2 respectively indicating the PHE and PAMR obeying the conventional empirical law. Below the CDW transition, S$_{xy4}$, C$_{xy4}$, and C$_{xy6}$ increase with the decreasing temperature and exhibit a strong field response with more contributions for oscillatory $\rho_{xy}$ and $\rho_{xx}$ at high fields. The temperature and field dependence of high-order oscillatory components (S$_{xy4}$, C$_{xy4}$, and C$_{xy6}$) suggests the strong enhancement of a unique coupling between anisotropy from lattice and electrons scattering for planar transport behavior.

Accompanying with CDW transition, it is observed that obvious multiple fluctuations emerge simultaneously. From lattice side, strong phononic fluctuations were revealed accompanying with lattice distortions\cite{Alaska}. From the electronic side, the electronic instability can host electronic fluctuations (including the nematic fluctuations) and result in electronic nematicity at lower temperatures with symmetry reduction\cite{Haoxiang}. Moreover, from magnetism side, for KV$_{3}$Sb$_{5}$ though in absence of long-range or short range magnetic order, the magnetic fluctuations due to observed orbital ordering would inherit the crystalline anisotropy of the kagome lattice which plays a crucial role in the magnetic properties especially for the appearance of time-reversal symmetry breaking. We find that the observed $\rho_{\parallel}<\rho_{\perp}$ (see Fig.2) is consistent with this picture since a magnetic field will suppress magnetic fluctuations in the field direction. In fact these fluctuations are correlated and can origin from same instability due to CDW transition indicating presentence of multiple orders in the system. These correlated fluctuations can be attributed to mediate the crystalline anisotropy and electron scatterings for planar  resistivity tensors to behave higher-order oscillatory components (as shown in Fig.4 (f)) revealing a unique interplay of lattice, electron and  magnetism in KV3Sb5.

In summary, prominent PHE and PAMR are observed in KV$_3$Sb$_5$. The applied field would drive higher-order oscillations for  both $\rho_{xy}$ and $\rho_{xx}$ below the CDW transition, resulting in the violation of the empirical law. This exotic phenomena suggest a strong and complicated coupling between lattice symmetry and  electron scatterings for planar transport properties. We expect that such unconventional PHE and PAMR are not only potential for next generation 3D chips, but also reveal a unique interplay between various degrees of freedom for materials with frustrated crystal structures, lattice distortions and anisotropic magnetic and fluctuations.

\begin{methods}
Single crystals of KV$_3$Sb$_5$  were grown via the self-flux method. K, V, and Sb were mixed by the ratio of 1:1:4.5 and sealed inside a Ta crucible. This Ta crucible was subsequently inside in  evacuated quartz tube. The mixture was heated to 1000 $^o$C, slowly cooled to 700 $^o$C, and finally decanted by a centrifuge.  Planar single crystals with typical dimensions of 2$\times$3$\times$0.3  mm$^3$ were harvested at the bottom of the crucible.

Crystal structure and elemental composition were confirmed by x-ray diffraction (XRD) and Energy Dispersive spectrometer(EDS). Sharp peaks in the XRD confirm high crystalline quality of the samples.

The Hall-bar devices were fabricated by exfoliating the single crystal of KV$_3$Sb$_5$ on to a SiO$_2$/Si substrate.  Electronic contacts were added via patterned mask with photolithography subsequent growth of Ti/Au(10nm/100nm) layers.  Electrical transport measurements were performed in a Physical Properties Measurement System (PPMS Dynacool, Quantum design). Direct current (DC) magnetization and alternating current (AC) susceptibility measurements were performed in a Magnetic Property Measuring System (MPMS Quantum design).

\end{methods}

\textbf{DATA AVAILABILITY}
\\The data supporting the findings of this study are available within the paper and in
the Supplementary Information, and also are available from the corresponding authors upon reasonable request.

\begin{figure}
  \centering
  \includegraphics[width=3.2in]{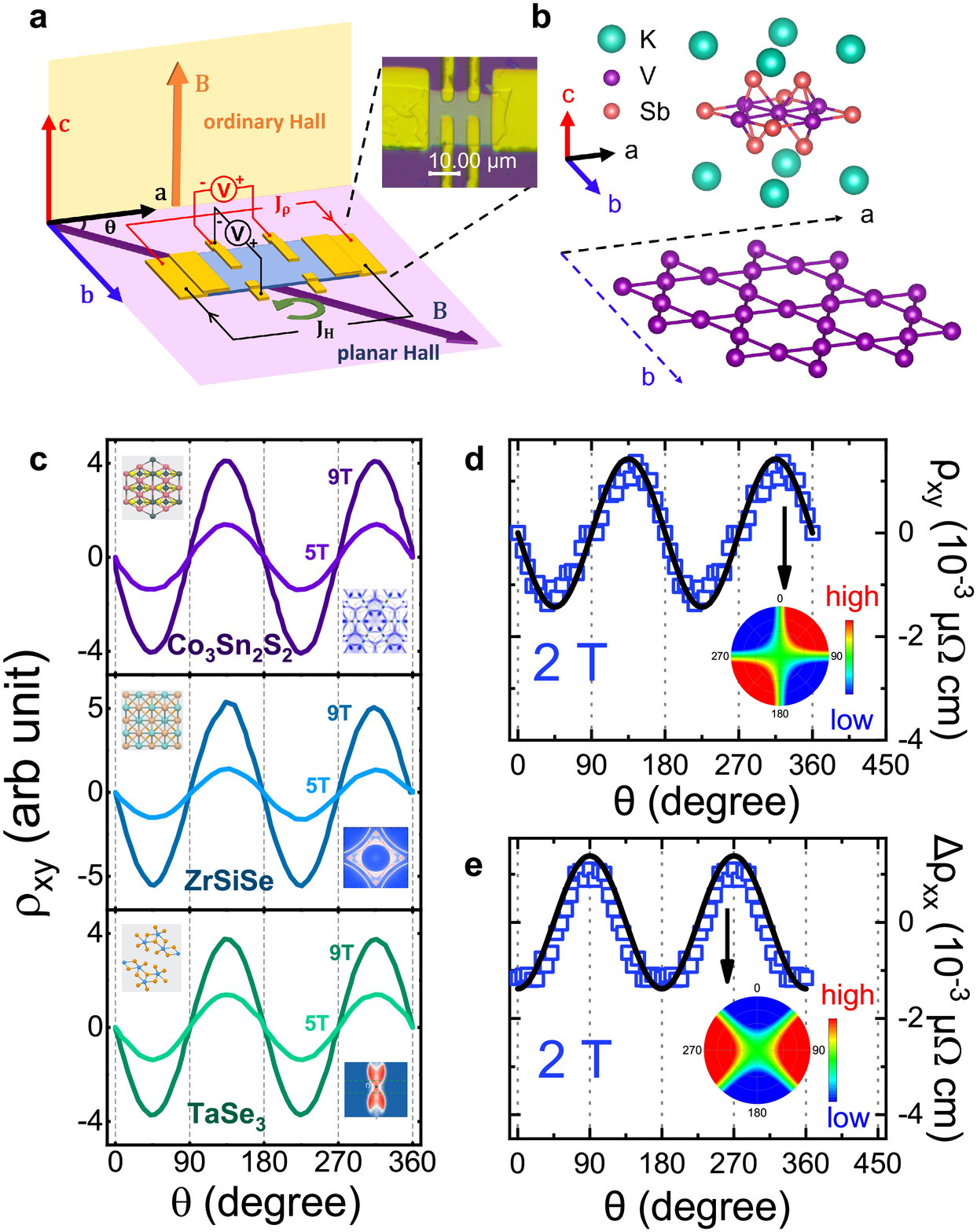}
  \caption{(a) Schematic of the geometry of the device in our measurements for the ordinary Hall, planar Hall, and longitudinal resistivity. Inset: the picture of our device. (b) Crystal structure of KV$_3$Sb$_5$.  $\theta$ is defined as the angle between the directions of current and in-plane applied field. For our device, the current is along $a$ axis. (c): The PHE in Co$_3$Sn$_2$S$_2$, ZrSiSe, and TaSe$_3$\cite{ShYY, Qiunan, ChenFC, Weib, YangXC, Gatti}. The insets: the crystal structure and Fermi surface in these three materials. (d) and (e): $\rho_{xy}(\theta)$ and $\rho_{xx}(\theta)$ with $\mu_0 H$=2 T at 10 K marked by blue open squares. The black solid lines indicate the $\rho_{xy}(\theta)$ and $\rho_{xx}(\theta)$  for conventional PHE and PAMR described by formula 1 and 2 respectively, which are also plotted in the insets exhibiting twofold symmetry by the color contours in polar coordinates where the $r$ axes are $\rho_{xy} (\theta)$ and $\rho_{xx} (\theta)$ respectively, and $\theta$ is the rotating angle defined in (a).
 }
  \label{fig:Fig1}
\end{figure}

\begin{figure}
  \centering
  \includegraphics[width=6in]{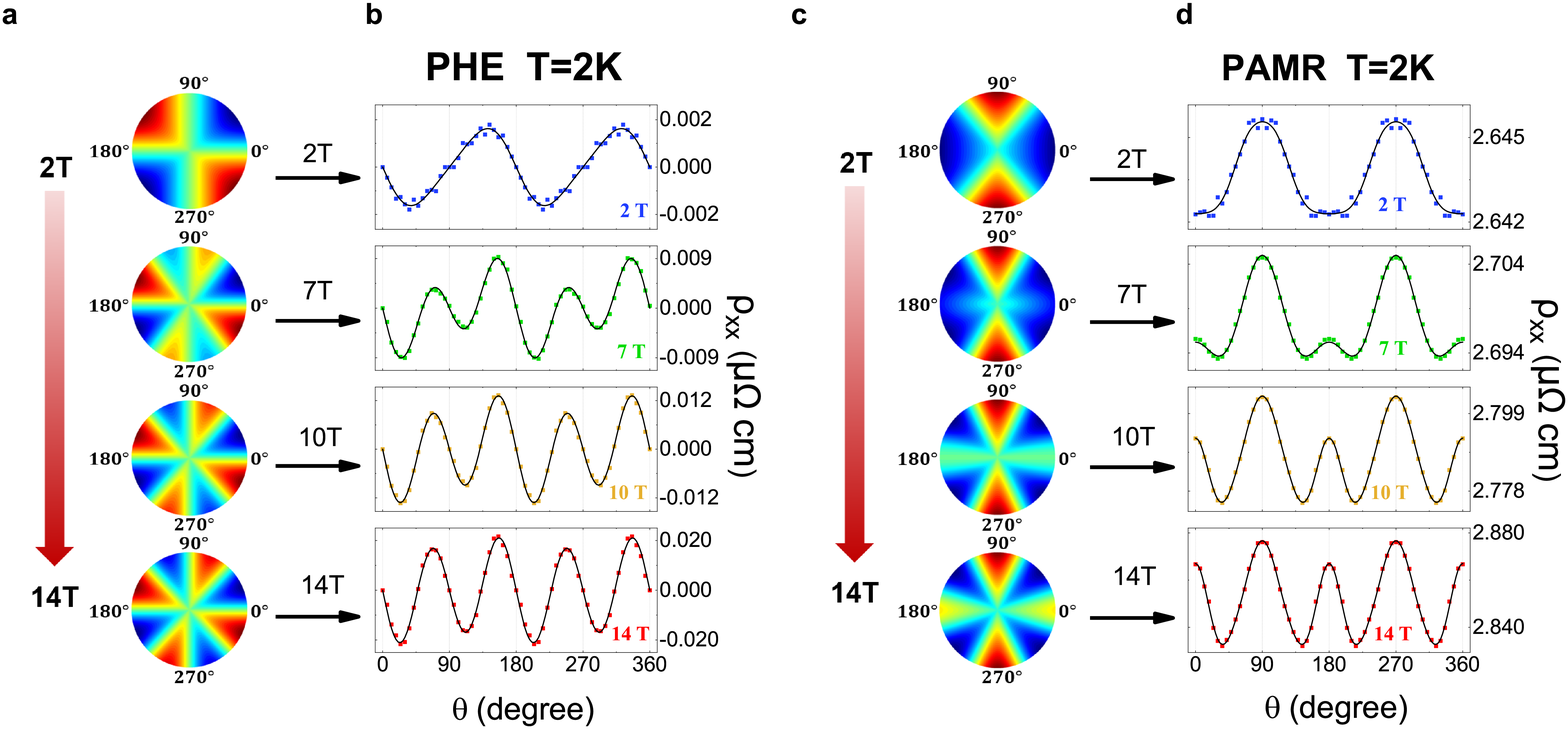}
  \caption{(a) and (c): The 2D color plots for PHE and PAMR at 2 K with $\mu_0 H$ = 2 T, 7 T, 10 T, and 14 T respectively in the polar coordinates where $r$ represents the applied field, $\theta$ is the rotating angle, the red region represents higher resistivity, and the blue region represents lower resistivity. (b) and (d): The related data of $\rho_{xy} (\theta)$ and $\rho_{xx} (\theta)$ from (a) and (c). The experimental data are marked by open symbols and the fitting curves by formulas 3 and 4 marked by solid lines.
 }
  \label{fig:Fig2}
\end{figure}

\begin{figure}
  \centering
  \includegraphics[width=3in]{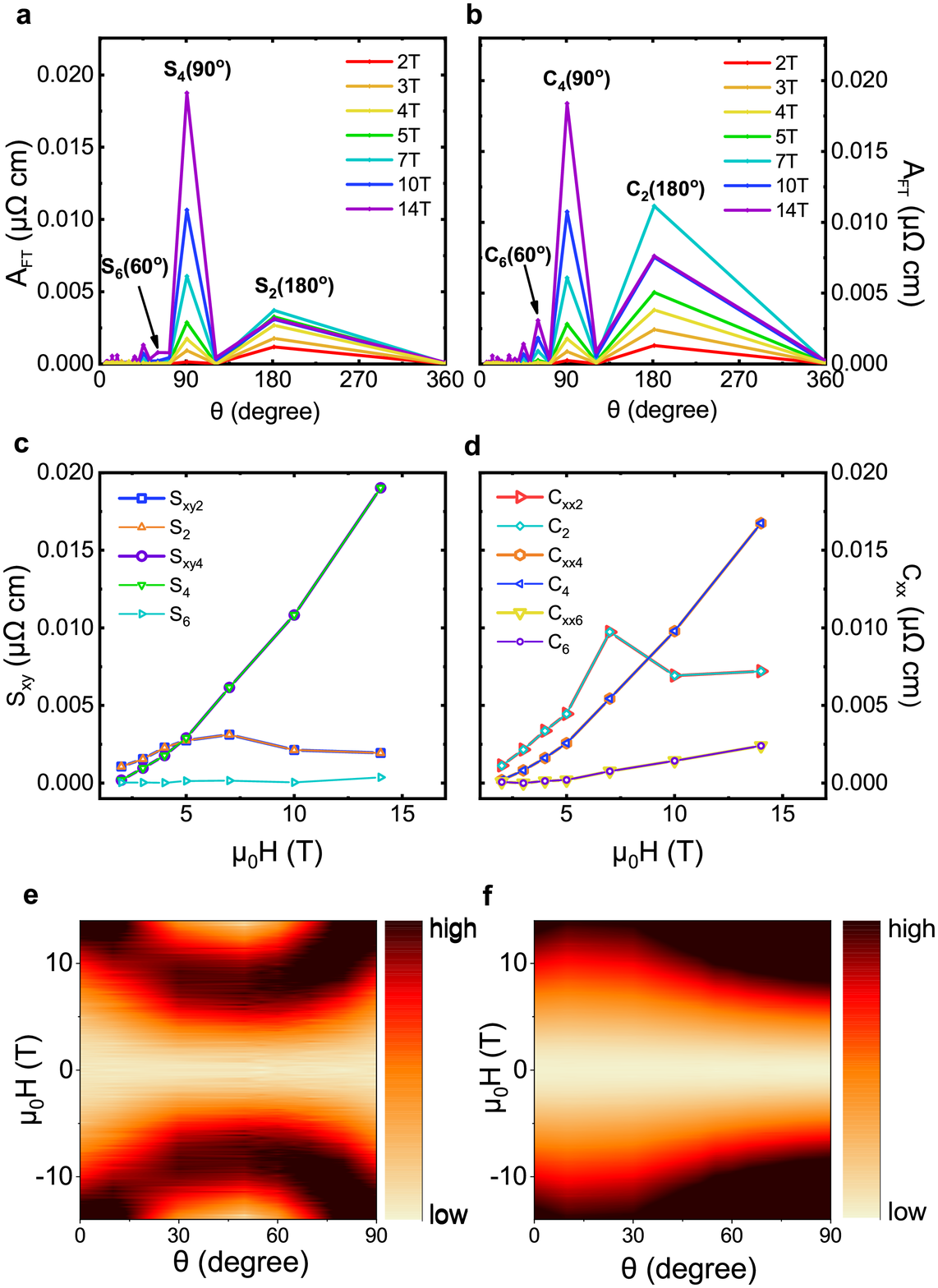}
  \caption{(a) and (b): The Fast Fourier Transform (FFT) spectrum of $\rho_{xy} (\theta)$ and $\rho_{xx} (\theta)$ at 2 K with $\mu_0 H$ = 2 T, 3 T, 4 T, 5 T, 7 T, 10 T, and 14 T. (c) and (d): The field dependence of high-order oscillatory coefficients for PHE and PAMR. $S_{2}$, $S_{4}$, $S_{6}$, $C_{2}$, $C_{4}$, and $C_{6}$ are  acquired from the amplitude of peaks of FFT spectrum of (a) and (b) . $S_{xy2}$, $S_{xy4}$,  $C_{xx2}$, $C_{xx4}$, and $C_{xx6}$ are acquired from the data fitting for the formulas (3) and (4). (e) and (f):  The color plots for field dependent  planar Hall and longitudinal resistivity $\rho_{xy}(\mu_0 H)$ and $\rho_{xx}(\mu_0 H)$ at the angles from 0$^{\circ}$ to 90$^{\circ}$ with the step of 10$^{\circ}$.
 }
  \label{fig:Fig3}
\end{figure}

\begin{figure}
  \centering
  \includegraphics[width=3in]{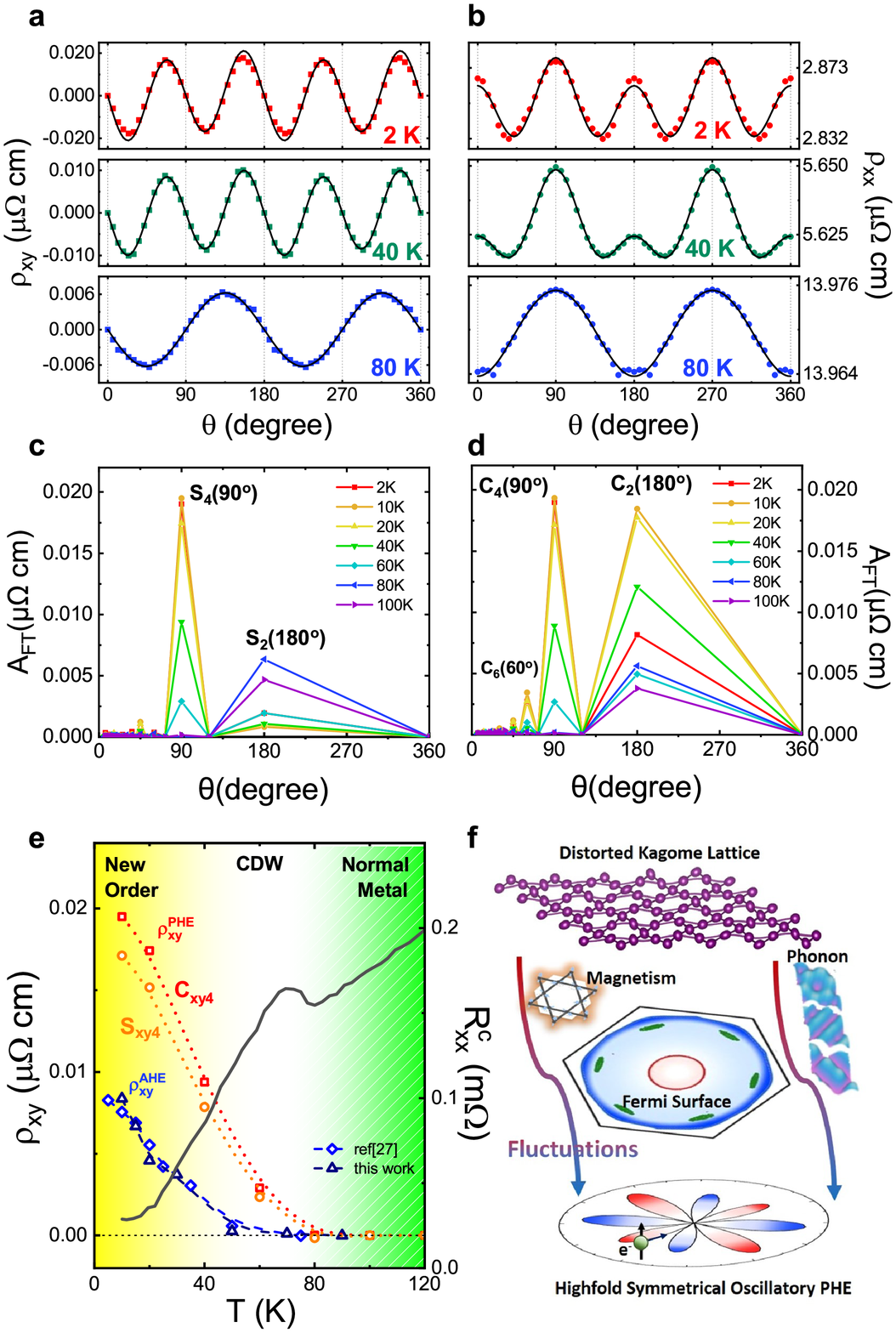}
  \caption{(a) and (b): The typical $\rho_{xy} (\theta)$ and $\rho_{xx} (\theta)$ with $\mu_0 H$ = 14 T at 2 K, 40 K, and 80 K. The fitting curves are marked by black solid lines.  (c) and (d): The FFT spectrum of $\rho_{xy} (\theta)$ and $\rho_{xx} (\theta)$ with $\mu_0 H$ = 14 T at 2 K, 10 K, 20 K, 40 K, 60 K, 80 K, and 100 K, respectively.  (e) : The temperature dependent anomalous Hall resistivity $\rho_{xy}^{AHE}$ for two samples are shown with the left axis. The data of black line are acquired from our measurements and the data of blue line are acquired from ref\cite{ShuoYingYang}. The temperature dependent \textit{S$_{xy4}$} and \textit{C$_{xx4}$} for $\rho_{xy}$ are also shown with the left axis. The temperature dependent resistance  along $c$ axis R$_{xx}^c$  is shown with the right axis with a prominent CDW anomaly around 78 K .
(f) The schematic for the coupling between distorted lattice and Fermi surface leading to  high symmetrical oscillatory PHE driven by an in-plane field. }
  \label{fig:Fig4}
\end{figure}


%


\begin{addendum}
 \item Work are supported by National Natural Science Foundation of China (NSFC) (Grants No.U213010013, 92165204,11904414, 12174454, and 11904417), Guangzhou Basic and Applied Basic Research Foundation (Grant No. 2022A1515010035,2021B1515020026), Guangzhou Basic and Applied Basic Research Foundation (Grant No. 202201011798), open research fund of Songshan Lake materials Laboratory 2021SLABFN11, National Key Research and Development Program of China (Grant No. 2019YFA0705702), OEMT-2021-PZ-02, and Physical Research Platform (PRP) in School of Physics, Sun Yat-sen University. The experiments reported were conducted on the Guangdong Provincial Key Laboratory of Magnetoelectric Physics and Devices (LaMPad).

 \item[Competing Interests] The authors declare that they have no competing interests.
  \item[Author contribution] L.L. and E.Y. carried on the single crystal growth, transport measurements, and contributed equally to this work. E.Y and B.W. performed the device fabrication. L.L. perform the numeric simulations for device. G.Y. performed transport measurements. B.S. conceived the ideas, contributed to most of the experiments and data processing. B.S., Z.Y., and M.W discussed and prepared the manuscript.
 \item[Correspondence] and requests for materials should be addressed to Bing Shen, Zhongbo Yan or Meng Wang.
\end{addendum}


\end{sloppypar}
\end{document}